\documentclass[11pt]{article}
\usepackage{graphicx}
\usepackage{amssymb}
\usepackage{enumerate}

\begin{document}

\title{Architecture for Community-scale Critical Infrastructure Coordination for Security and Resilience}
\author{James~Christopher~Foreman\\Purdue University \\401 N. Grant St. West Lafayette, IN 47907}
\date{5 Jul 2016}

\maketitle

\section*{Abstract}
\textbf{Our Critical Infrastructure (CI) systems are, by definition, critical to the safe and proper functioning of society. Nearly all of these systems utilize industrial Process Control Systems (PCS) to provide clean water, reliable electricity, critical manufacturing, and many other services within our communities -- yet most of these PCS incorporate very little cyber-security countermeasures. Cyber-attacks on CI are becoming an attractive target. While many vendor solutions are starting to be deployed at CI sites, these solutions are largely based on network monitoring for intrusion detection. As such, they are not process-aware, nor do they account for interdependencies among other CI sites in their community. What is proposed is an architecture for coordinating all CI within a community, which defines characteristics to enhance its integration, its resilience to failure and attack, and its ultimate acceptance by CI operators. }

\section{Introduction}
Our critical infrastructure systems provide the supporting services necessary for safe and productive living within our society. They are our electrical power, treated water, natural gas utilities as well as roads, dams, hospitals, and telecommunications. The White House and Department of Homeland Security have defined critical infrastructure and established the priority through ongoing presidential directives and policy through at least the last three presidents \cite{clinton1996executive,bush2003homeland,house2013presidential}. As such infrastructure is critical to the safe functioning of communities and society, ensuring the secure and reliable operation of critical infrastructure becomes a paramount concern and goal.

\subsection{Problem Description}
A large majority of our critical infrastructure depends on Process Control Systems (PCS) to accomplish their mission, such as operating dams, performing water treatment, generating electrical power, controlling critical manufacturing, etc. Included within PCS are Supervisory Control and Data Acquisition (SCADA), Programmable Logic Controllers (PLC), and Distributed Control Systems (DCS). These systems have generally been late to implement significant cyber security countermeasures, and have thus recently been the subject of attacks \cite{cardenas2011attacks,shackelford2015securing}. In the past, PCS were built with custom systems and air-gapped from the Internet, serving only the sensors and actuators of their respective processes. Starting around the turn of the 21st century, the trends to replace these systems with Commercial-Off-The-Shelf (COTS) industrial hardware and mainstream operating systems, the interconnection with corporate WANs, and other modernization techniques has created a new and significant attack surface \cite{schukat2014securing,foreman2015identifying}, which is expanding daily as new vulnerabilities are discovered and new attacks developed. 

Since these processes are so critical to the continued functioning of society, they present a very attractive target to malicious cyber attacks, whether the attackers are nation-state terrorists, or simply opportunistic crackers. It is inevitable that such cyber attacks on our critical infrastructure will rise significantly in the near term and that the current lack of preparedness will result in significant disruption to our society \cite{sanger2012rise}. Further exasperating this security problem is the fact that most critical infrastructure processes themselves rely on other critical infrastructure processes, e.g., water delivery relies on electrical power. Understanding and controlling this interdependence is a large problem in itself \cite{rinaldi2001identifying,dudenhoeffer2006cims}, which greatly complicates the building of coordinated solutions and sets the stage for significant cascading failures.

\subsection{Needed Approach}
What is needed is a community-scale system to provide anomaly/intrusion detection, mitigation planning, and a coordinated response. This will ensure secure critical infrastructure resilience through situation awareness and facilitate mitigating responses that build on the infrastructure interdependencies as an asset, versus the chaos of handling multiple infrastructure as a complexity. To meet this need, architecture is proposed to integrate multiple critical infrastructure PCS to provide a coordinating link for centralized management.

\section{Previous Work}
There have been several efforts related to the securing of individual critical infrastructure with a cyber system. Most of these in practice employ the approach of adding COTS security solutions from the field of Information Technology (IT), e.g., firewalls. While these represent a best practices approach to basic computer network security, they are far too ineffective as a primary security layer for critical infrastructure. One reason is that current IT approaches do not take into account the unique software, hardware, and protocols used in PCS, though some vendors have released products to address this, e.g., Tofino\textregistered \cite{kang2014high}. Another reason is that once an intrusion is detected, the PCS network must remain open to continue the critical infrastructure process, as a shutdown would be too disruptive to society. Therefore, mitigation and adaptation approaches are best and the focus of this survey will be on such intelligent approaches.  

\subsection{Individual Critical Infrastructure}
Research efforts have been devoted to the development of intrusion detection techniques that work on the network layer of PCS \cite{cheung2007using,dussel2009cyber}. Such research has resulted in software models specific to PCS protocols and traffic patterns. This type of work extends the traditional IT approaches to the PCS domain, but still lacks in mitigating approaches after attack, and does not engage the physical critical infrastructure process itself; although Cheung et al. \cite{cheung2007using} discuss using the network model as a detector for anomalous traffic. Other approaches that consider the process under control \cite{genge2013data,linda2009neural} have additional capabilities in that they can detect when the infrastructure process is behaving abnormally. These capture attacks even in scenarios of normal network traffic and proper authentication; some solutions also include scenarios of operator error and equipment malfunction. Many other efforts exist and fall into one or both of these categories of solutions that focus on system configurations specific to PCS, and those that focus on the validity of the process being controlled, e.g., SOCCA \cite{zonouz2014socca} provides situational awareness and contingency analysis for the power grid utilizing grid state estimation techniques to provide security. These show the best merit moving forward, but do not include coordination with other critical infrastructure to achieve community-scale situational awareness and event management. For these, we need a community-scale approach, and architecture of how to connect these.

\subsection{Coordination of Multiple Critical Infrastructure in a Community}
Again, the focus is on intelligent approaches that develop mitigating and adapting responses to malicious or anomalous activities, now applied simultaneously among multiple critical infrastructure taking advantage of interdependencies. In this scenario, a bigger net is cast on the detection of cyber attacks within communities by monitoring multiple PCS, better situational awareness is achieved, and calculated responses can consider the effects across the domain of the whole community.

Branscomb et al. \cite{branscomb2006sustainable} identify communities (cities) as a mass infrastructure critical to sustaining society and make the case for the need of cooperation among infrastructure providers for catastrophic risk reduction, noting vulnerabilities and potential cascading failures from interdependency. Auerswald et al. \cite{auerswald2006seeds} further supports this need with respect to the problem of balancing the efficiency of critical infrastructure, which is typically owned and operated by the private sector, with mitigating those critical infrastructure vulnerabilities that greatly impact the community at large (public). Both of these \cite{branscomb2006sustainable,auerswald2006seeds} introduce policy approaches that can lay the foundation for building a cooperative solution, but do not explicitly define any system architecture for such a solution.

Other efforts have proposed methods to harness existing efforts and build partnerships for community critical infrastructure protection between public and private entities \cite{o2013unraveling}. Understanding the complexities of disaster management \cite{may2013addressing}, and developing disaster-resilient communities \cite{chen2013public} have also been explored. The proposed architecture needs to address these policy and partnership aspects while providing a foundation for common connectedness among critical infrastructure systems. This facilitates the acceptance of community-based solutions so they can be attempted and evaluated.

Murray and Grubesic \cite{murray2012critical} recognize the vulnerability conundrum in that reducing the worst-case vulnerabilities does not necessarily ensure the best allocation of protective resources. This demonstrates the need for intelligent coordination in a solution. A particular critical infrastructure in a community may mitigate their worst-case attack while leaving other critical infrastructure more vulnerable or adversely affected by such protective actions, e.g., the rerouting of power during power infrastructure cyber-attacks impacting the delivery of community drinking water. Additionally, the benefits of multiple critical infrastructure working together are not enjoyed in individual solutions. Murray and Grubesic \cite{murray2012critical} further expand this to discuss multi-community protective planning for protection from attacks.

Yusta et al. \cite{yusta2011methodologies} provides an extensive review and categorization of the state-of-the-art, c. 2011, in critical infrastructure protection software, which includes numerous and specific cyber-security approaches. These approaches focus on vulnerability, risk, and interdependency identification and assessment to realize risk management systems by employing modeling and simulation. In all of these, the goal is to understand the problem of critical infrastructure cyber-security in the best detail possible, but the architecture for mitigation and the advancement to a coordinating solution is still not realized. 

\section{The Proposed Architecture}
The goal of this proposed architecture is to enhance PCS utilized in critical infrastructure to realize a Community-scale Critical Infrastructure Coordination (CCIC). This will be employed in communities for the security and resiliency of critical infrastructure and thus, the normal functioning of society. 

\subsection{Definition of Key Characteristics}
First, it is important to define a few characteristics that are key to a successful architecture for the CCIC enhancement.
\begin{enumerate}[1.]
\item It must be relatively simple in order to be integrated within a wide range of PCS vendors, configurations, and vintages of legacy systems.
\begin{enumerate}[a.]
\item This integration must also be simple for operators to use effectively, minimizing training and maximizing acceptance.
\item Key to maximizing acceptance is the ability to guarantee that the critical infrastructure process not be halted or negatively impacted by the CCIC, even if the CCIC is maliciously compromised or experiences a failure mode. 
\end{enumerate}
\item It must incorporate learning to fit to the unique operating states of a specific critical infrastructure process to minimize the impact of intensive installation effort and cost at every site, which would significantly delay deployment of the CCIC.
\begin{enumerate}[a.]
\item Learning must be automated and data driven to capture the unique aspects of the process, but must avoid being too opaque, i.e., black box, which would further hinder diagnosis, monitoring, and acceptance by operators.
\item This learning must also adapt to changing process conditions as the actual equipment ages, wears, or otherwise drifts out of calibration. 
\end{enumerate}
\item It must be hierarchical in order to form links with peer critical infrastructure and uplinks to higher-layer management. 
\begin{enumerate}[a.]
\item These links should be bi-directional to provide both situation awareness and mitigating responses.
\item The CCIC should maintain some functionality at a local layer during the loss of any or all of these links.
\end{enumerate}
\end{enumerate}

\subsection{Defining the Local Hierarchical Layer}
Secondly, the hierarchical layers need to be defined. At the first and local layer, i.e., at a single critical infrastructure site, the architecture is integrated within the existing PCS as illustrated in Fig. 1.
\begin{figure}[htb]
\begin{center}
\includegraphics[]{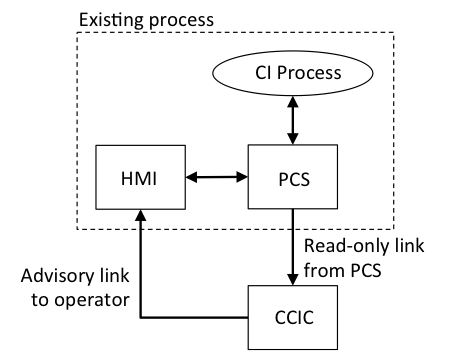}
\caption{Integrating at the local site, hierarchical layer 1.}
\end{center}
\end{figure}
In Fig. 1, the architecture satisfies characteristic 1 by being positioned outside the existing Critical Infrastructure (CI) process. Most modern PCS provide some interface for reading process variables and can specify a read-only link, preferably through hardware definition. The read-only link to the PCS ensures that under no circumstances can the CCIC, either layer 1 or 2, write back (send) control commands to the PCS, intentional or otherwise, satisfying characteristic 1b. These process variables are then monitored at the local layer with respect to a learned model to classify operating states and detect anomalous behavior. Variables that exhibit this behavior are sent to a rule-based expert system, which provides reports and recommendations to the operator via the Human Machine Interface (HMI). The variables may be simply reported, or the detection of anomalous behavior of certain variables may direct the operator to take recommended actions. The final decisions still go through a human, using the already familiar and existing HMI that is experienced in the proper operation of the critical infrastructure, thus satisfying characteristic 1a. A block diagram of this activity is described in Fig. 2.
\begin{figure}[htb]
\begin{center}
\includegraphics[]{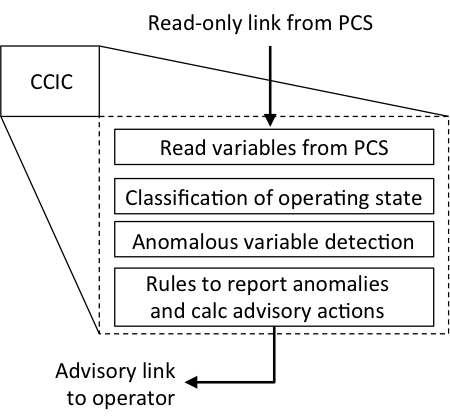}
\caption{Activities within hierarchical layer 1.}
\end{center}
\end{figure}
The architecture needed to perform the activities of Fig. 2 requires a classification engine and an expert system, and is illustrated in Fig. 3. Together, these achieve the characteristics in 2 including 2a and 2b. Artificial Neural Networks (ANN) provide an excellent option for automated learning of patterns from data, and modern computing systems can easily execute these in near real time for a critical infrastructure process. However, a significant problem with ANN is its black-box approach. PCS depend on being completely determined, or at least nearly so, in operation. An ANN removes this certainty, especially when dealing with a nontrivial number $(>>100)$ of input and output variables. One approach is to not depend on the ANN for overall classification of the process, but to modularize different control sections, e.g., PID loops, with a small $(<<10)$ ANN for each of these. This simplifies the ANN allowing its operation to be more completely understood due to its reduced dimension. The use of smaller, modular ANNs, as well as safeguards on classification errors can be mitigated with Rule-Based Expert Systems (RBES). These are intended to provide the glue logic and other bounds checking prior to incorporation into an advisory output to the operator. \cite{foreman2008architecture}
\begin{figure}[htb]
\begin{center}
\includegraphics[]{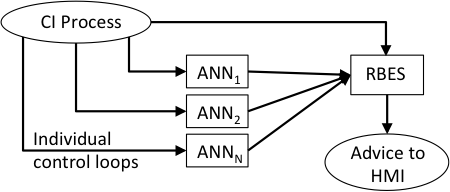}
\caption{Architecture of hierarchical layer 1.}
\end{center}
\end{figure}
As shown in Fig. 3, small ANNs classify the operating state of individual control loops from various sub processes against known, valid operating states. When an anomaly is detected, this is sent to the RBES, which determines an advisory message to send to the operator for awareness, possibly suggesting corrective actions. For changes in equipment due to age, wear, etc. Ð this can be handled through partial retraining of the ANN periodically, e.g., replacing 20\% of the original training data set with recently collected, non-anomalous process data. Similar work has been done in the utilization of sensor validation networks, e.g., \cite{upadhyaya1992application}, such as virtual sensors and detecting drift in sensor calibration. 

\subsection{Defining the Community Hierarchical Layer}
While the previous section described functionality at the local layer, community-scale solutions need to occur at a hierarchical layer above this, i.e., the community layer, 2. The community layer coordinates the local layer at various critical infrastructure, e.g., power generation, distribution, water treatment and delivery, etc., with each other at a centralized point for community planning. This centralized point can be a city council, mayorÕs office, or center for catastrophic event response. This architecture is illustrated in Figs. 4 and 5.
\begin{figure}[htb]
\begin{center}
\includegraphics[]{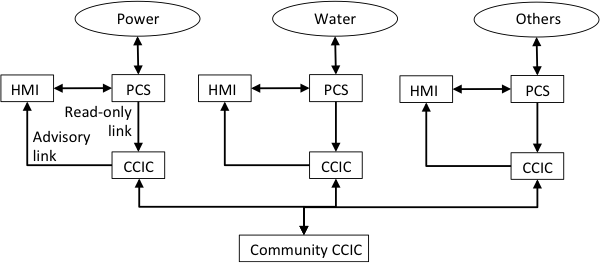}
\caption{Overall architecture of the community layer, hierarchical layer 2.}
\end{center}
\end{figure}
In the diagram shown in Fig. 4, the local layer for each individual infrastructure is interconnected to a community layer through bi-directional links. Characteristic 3b is achieved by definition, as defined in the architecture at the local layer, and the extension to the community layer enables characteristic 3a. It is important to note that the individual sites may coordinate with each other without the community layer to handle some interdependencies and situational awareness, e.g., the water company can be alerted to cyber-attacks on the power system that may affect certain water delivery pumps.
\begin{figure}[htb]
\begin{center}
\includegraphics[]{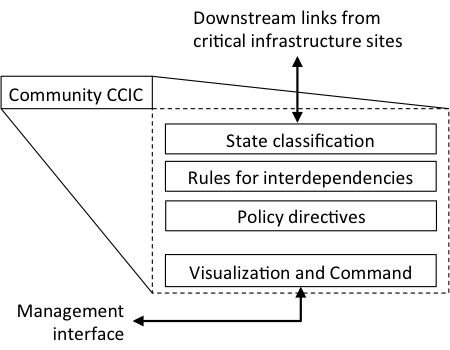}
\caption{Activities of hierarchical layer 2.}
\end{center}
\end{figure}
The activities depicted in Fig. 5 can be performed with the same ANN and RBES hybrid as employed at the local layer, Fig. 3. The community layer monitors status variables from each critical infrastructure site similarly to control loops with ANNs. These become classifications of operating state and detected anomalies, which are then fed to a RBES. The RBES handles interdependencies between infrastructure based on expert knowledge and applies policy directives to calculate responses to anomalous behavior and catastrophic events. 

Thus, the classifications provide for a community-scale situational awareness through a visualization interface, and the rule sets handle mitigating responses as an intelligent advisor, similar to the advisory output made available to the operator through the HMI at the local layer. The command interface allows manual input of actions to coordinate critical infrastructure or through automation for certain actions as appropriate. This takes all critical infrastructure sites into consideration to achieve an overall optimal solution versus competing local optimizations, solving the vulnerability conundrum \cite{murray2012critical} and further satisfying characteristic 3.

\subsection{Data Centric Aspects of the Architecture}
The architecture is data-centric through its utilization of process variables (data) at the local layer for training and classification of operating state. The large numbers of sensory inputs, typically 100s to 1000s for critical infrastructure processes, ensure a sufficient level of dimensionality to accurately classify operating state, even in the presence of outliers. These outliers, which are process variables outside of their normal range for a given operating state, are used to determine anomalous behavior. The RBES can recall pre-programmed advisory statements to send to the operator for given outliers. For example, a pumpÕs rotational speed is much higher than expected for a given flow, therefore the pump may be worn or the flow sensor may be out of calibration. If examination of other process variables eliminates both of these cases, a possible conclusion would be an intruder spoofing the process variables in the PCS. Most critical infrastructure PCS in operation today have years of real time data, and only operate in a finite state-space, providing a data rich environment for model training and classification. These aspects are illustrated in Fig. 6.
\begin{figure}[htb]
\begin{center}
\includegraphics[]{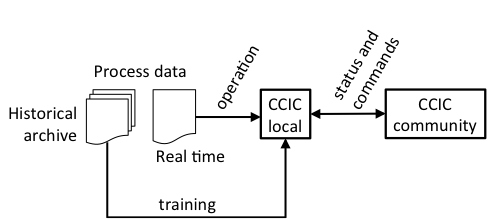}
\caption{Data centric aspects of the architecture.}
\end{center}
\end{figure}

\subsection{Event Driven Aspects of the Architecture}
The architecture is also event driven through its actions, advisory or automated, based on anomalous activity, i.e. anomalous classification of process variables and detection of outliers. The event-driven advisory feedback to local layer operators drives mitigating responses at the local level through the operator, i.e., human in the loop. At the community layer, events at other critical infrastructure sites may influence a given site, and events created by the community layer managersÕ actions flow downstream to manually guide local layer sites. At the local layer, classification of anomalous process data triggers the action of a mitigating response, e.g., advisory output. In the previous example on data centric aspects section, the events include operator verification of the process and technician validation of equipment. The actions performed at the local layer become events that trigger subsequent actions at the community layer. In the previous example this might include: preparing other infrastructure dependent on this process of potential outages; alerting the community of disruption of service; or alerting all sites and deploying network countermeasures if a cyber-attack is determined. These aspects are illustrated in Fig. 7.
\begin{figure}[htb]
\begin{center}
\includegraphics[]{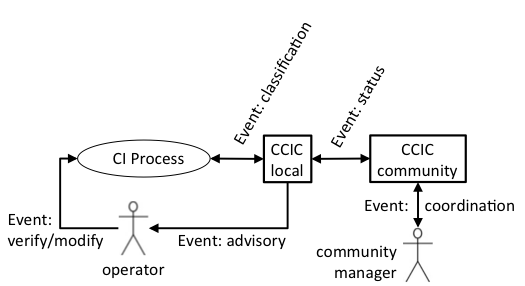}
\caption{Event driven aspects of the architecture.}
\end{center}
\end{figure}

\subsection{Use Cases of the Architecture}
In Fig. 8, a general use case diagram is illustrated. The two primary human actors are the process control system operator at the local critical infrastructure site, and a city council operator at a community-planning site. Communications between these two may take place externally, e.g., discussion via telephone, or through the architecture itself. The city council operator sees all infrastructure within the community and can thus influence the local layer through the bi-directional coordination links. This coordinating information is taken into the local model, along with the local process data, to build the advisory outputs to the operator.
\begin{figure}[htb]
\begin{center}
\includegraphics[]{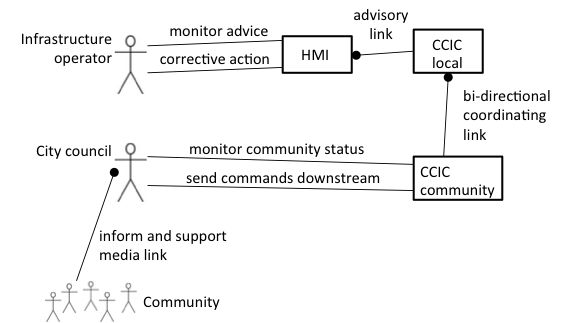}
\caption{Use cases of the architecture.}
\end{center}
\end{figure}
The infrastructure operator receives advice from the local layer. This advice is constructed to provide straightforward and meaningful information on anomalies detected, pointing to specific process points or equipment, which can be investigated by local personnel. The advice may also include suggested corrective actions derived from the RBES with rules set by local engineering personnel. As this advisory/response process occurs, the status of the infrastructure process is sent upstream to the community layer.

At the community layer, a city council operator keeps community management personnel informed, which in turn alerts the community as needed. The city council operator can take mitigating actions. Some sample scenarios follow.

Sample scenario 1: Cyber-attack detected at water treatment plant
\begin{itemize}
\item Local layer
\begin{itemize}
\item Detect attack through anomalous sensor readings
\item Isolate water treatment plant from Internet
\item Validate suspect sensors and control logic
\item Place in manual control and initiate network cyber-security countermeasures
\end{itemize}
\item Community layer
\begin{itemize}
\item Isolate other critical infrastructure networks as precaution
\item Check dependencies on other critical infrastructure, i.e., clean water may become unavailable, plan accordingly
\item Determine community impact and affected areas to verify water is safe
\end{itemize}
\end{itemize}
Sample scenario 2: Catastrophic failure of power grid, e.g., storm or earthquake
\begin{itemize}
\item Local layer
\begin{itemize}
\item Detect failure through alarms and sensor data
\item Determine scope and attempt to localize faults to protect the grid
\item Increase monitoring of remaining power grid sections
\end{itemize}
\item Community layer
\begin{itemize}
\item Check dependencies on other critical infrastructure, i.e., water delivery and telecommunications disruptions from equipment powered by the affected section of the power grid, re-route water distribution and communications
\item Determine community impact, initiate water rationing, limit communications bandwidth
\end{itemize}
\end{itemize}

\section{Conclusions}
The architecture for community-scale coordination of critical infrastructure has been designed to provide an intelligent approach to securing the resilience of critical infrastructure process control systems. Through training and expert knowledge, a process-aware model is created at the local layer to detect and advise on local process anomalies. By up-linking these local models, a community-scale approach is realized. All critical infrastructure in a community can share in detection of anomalous behavior, and their mitigating responses can be coordinated for the optimal solution for the whole community. Characteristics achieved are a simplified implementation through automated training on process data, while modularizing this training and applying rule-based logic minimizes the black box approach that is unfavorable in process control systems. Further enhancing acceptance are failure modes isolated from the process at the local layer through a read only interface, and the provision of only advisory outputs with human (operator) in the loop validation.

\bibliographystyle{unsrt}
\bibliography{CCIC-Arch}

\end{document}